
\input harvmac
\input epsf

\Title{BUHEP-95-21}
{\vbox{\centerline{Limits on Pseudoscalar Bosons}\vskip 12pt
\centerline{From Rare $Z$ Decays at LEP}}}

\centerline{Gautam Rupak\foot{E-mail: grupak@buphy.bu.edu.\ \  Current
address:
Physics Department, University of Washington, Seattle WA 98195} and
Elizabeth H. Simmons\foot{ E-mail: simmons@bu.edu}}
\bigskip
\centerline{Department of Physics}
\centerline{Boston University}
\centerline{590 Commonwealth Avenue}
\centerline{Boston, MA 02215}
\vskip .3in

Current LEP data on rare $Z$ decays constrain nonminimal technicolor
models that contain light neutral pseudo-Nambu-Goldstone bosons
($P^a$).  We discuss the production and decay of such particles, and
show how LEP data on $\gamma\gamma\gamma$, $\gamma + E\!\!\!/ $ and
$\gamma+$hadrons constrain the size of the technicolor gauge group and
the strength of the $Z\gamma P^a$ coupling.  The limits are then
applied to several specific technicolor scenarios.

\Date{7-95}

\def\pas{P^a}
\def\pa{$P^a$}
\def\mp{M_P}

\def\lta{\ \hbox{\raise.55ex\hbox{$<$}} \!\!\!\!\!
\hbox{\raise-.5ex\hbox{$\sim$}}\ }
\def\gta{\ \hbox{\raise.55ex\hbox{$>$}} \!\!\!\!\!
\hbox{\raise-.5ex\hbox{$\sim$}}\ }


\nref\manran{A. Manohar and L. Randall, Phys. Lett. {\bf B246} (1990)
537}  \nref\ransim{L. Randall and E.H. Simmons, Nucl. Phys. {\bf B380}
(1992) 3} \nref\lubicz{V. Lubicz, Nucl. Phys. {\bf B404} (1993) 559}
\nref\grdel{DELPHI Collaboration, P. Abreu {\it et al.}, Phys. Lett.
{\bf B327} (1994) 386}
\nref\grlthrephot{L3 Collaboration, Measurement of Energetic
Single-Photon Production at LEP, CERN-PPE/94-216 (1994)}
\nref\gropalphot{OPAL Collaboration, Measurement of Single Photon
Production in $e^+ e^-$ Collisions near the $Z^0$ Resonance,
CERN-PPE/94-105 (1994)}
\nref\grlthrehadr{L3 Collaboration, O. Adriani {\it et al.},
Phys. Lett. {\bf B292} (1992) 472}
\nref\pff{S. Dimopoulos, S. Raby and G.L. Kane, Nucl.
 Phys. {\bf B182} (1981) 77 \semi J. Ellis {\it et al.}, Nucl. Phys.
{\bf B182} (1981) 529
 \semi B. Holdom, Phys. Rev. {\bf D24} (1981) 1441}
\nref\esetc{S. Dimopoulos and L. Susskind, Nucl. Phys. {\bf B155}
(1979) 237 \semi
E.Eichten and K. Lane, Phys. Lett. {\bf B90} (1980) 125}
\nref\ranrius{See, e.g., L. Randall and N. Rius, Phys. Lett. {\bf
B309} (1993) 365 }
\nref\bench{E. Farhi and L. Susskind, Phys. Rep. {\bf 74}
(1981) 277}
\nref\appter{T. Appelquist and J. Terning, Phys. Lett.
{\bf B315} (1993) 139}
\nref\precision{
 M. Golden and L. Randall, Nucl. Phys. {\bf B361} (1991) 3 \semi
 M. E. Peskin and T. Takeuchi, Phys. Rev. Lett. {\bf 65} (1990) 964
\semi
 B. Holdom and J. Terning, Phys. Lett. {\bf B247} (1990) 88 \semi
 W. J. Marciano and J. L. Rosner, Phys. Rev. Lett. {\bf 65} (1990) 2963
\semi D. Kennedy and P. Langacker, Phys. Rev. Lett. {\bf 65} (1990)
2967; Phys. Rev. {\bf D44} (1991) 1591 }
\nref\eichlan{K. Lane and E. Eichten, Phys. Lett. {\bf B222} (1989)
274}
\nref\lanram{K. Lane and M.V. Ramana, Phys. Rev. {\bf D44} (1991)
2678}
\nref\lane{K. Lane, Color-Singlet Technipions at
the Tevatron, BUHEP-95-23, hep-ph9507289}
\nref\lubicztwo{V. Lubicz
and P. Santorelli, Production of Neutral Pseudo-Goldstone bosons at
LEPII and NLC in multiscale walking technicolor models, BUHEP-95-16,
DSF 21/95, hep-ph/9505336}

\newsec{Introduction}

Data from experiments at the Z pole
provide valuable information about models of electroweak symmetry
breaking.  In particular, LEP now probes rare Z decays
with branching ratios at the $10^{-5} - 10^{-6}$ level.  This allows
the possibility of detecting non-standard particles that couple very
weakly to the $Z$.

Many non-minimal technicolor models include electrically neutral
pseudo-Nambu-Goldstone bosons that couple anomalously to the
photon and $Z$.  These non-standard pseudoscalar particles (\pa) can be
produced at LEP through the process $Z \to
\gamma\pas$ \manran.  If the pseudoscalar's mass is less than about 65
GeV, the branching fraction can be of order $10^{-5}$.  A dramatic
feature of these rare $Z$ decays is the fact that the photon energy is
uniquely fixed by the scalar's mass. Previous investigations
\ransim\lubicz\ of the production and decay modes of these neutral
pseudoscalars have shown that all decays yield distinctive signatures
for which the LEP experiments can search.  Among the most
interesting are events with three final state photons, because neither
composite nor supersymmetric models should yield a visible
signal in this channel \ransim .

This paper evaluates the limits that LEP data on rare $Z$
decays \grdel\grlthrephot\gropalphot\grlthrehadr\ place on
non-minimal technicolor models.  We begin by  discussing the rate of
\pa\ production and the available decay modes.  Next, we explore the
kinematics of the final states for which experiments can search.
We study the limits on each final state and express them in
terms of bounds on the size of the technicolor gauge group and the
strength of the  $Z\gamma\pas$ coupling.  Finally, we apply our
bounds to several specific technicolor scenarios.

\newsec{Production and Decay of \pa\ }

At $\sqrt{s} = M_Z$, the primary\foot{The
sub-dominant processes $Z \rightarrow Z^* \pas
\rightarrow f \bar f \pas$ and $Z \to b \overline{b} \to b
\overline{b} \pas$ are discussed in \ransim \lubicz .}  production mode
for a neutral pseudo-Nambu-Goldstone boson (PNGB) of mass less than
$M_Z$ is the two-body process $Z \to \gamma \pas$.  For
technifermions in the fundamental representation of
technicolor, a PNGB couples to a pair of gauge bosons
${G_1,G_2}$ with charges ${g_1,g_2}$, momenta ${k_1,k_2}$ and
polarizations ${\epsilon_1,\epsilon_2}$ just as a pion couples to a
pair of photons
\pff :
\eqn\amp{N_{TC}{\cal A}_{G_1G_2} {g_1 g_2 \over{2 \pi^2 f}}
\epsilon_{\mu\nu\lambda\sigma} k_1^\mu k_2^\nu \epsilon_1^\lambda
\epsilon_2^\sigma .}
Here ${\cal A}_{G_1G_2}$ is the anomaly factor for the axial current $T^a$
associated with \pa\
\eqn\cala{4 {\cal A}_{G_1G_2} =
{\rm Tr}\left[ T^a (T^1 T^2 + T^2 T^1)_L\right] +
{\rm Tr}\left[T^a (T^1 T^2 + T^2 T^1)_R\right]}
$T^1$ and $T^2$ are the generators associated with the gauge bosons,
$L$ and $R$ refer to left-- and right--handed technifermions, and $f$
is the PNGB decay constant. In particular, the anomaly factor relevant
for $Z \to \gamma \pas$ is
\eqn\calb{{\cal A}_{Z\gamma} =
{1\over2} {\rm Tr} \left[T^a (T_{3L} + T_{3R} - 2 Q \sin^2\theta_W )
Q\right].}
so that if right handed technifermions are weak singlets,
a PNGB with custodial isospin 0 (1,2) has
${\cal A}_{\gamma Z}$ proportional to $\sin^2 \theta_W$ or
$1 - 2\sin^2 \theta_W$ ($1-4 \sin^2 \theta_W$, $1-2
\sin^2\theta_W$)\ \manran.

The $Z$--boson decays to $\gamma \pas$ with width \manran
\eqn\zdecay{\Gamma_{Z \rightarrow \gamma \pas}\
=\ 2.3 \times 10^{-5} {\rm GeV} \left({123 \over f} \right)^2 (N_{TC}
{\cal A}_{Z\gamma})^2 \left({ M_Z^2-\mp^2 \over M_Z^2}\right)^3}
where $M_Z$ and $\mp$ are the $Z$ and scalar masses,  It is
reasonable to expect a branching ratio of order $10^{-5}$, which is
large enough to be visible in current LEP experiments.

Once produced, the PNGB has many possible decay paths \ransim :

\medskip

\item{$\bullet$} In models where \pa\ only decays to electroweak gauge
bosons, the dominant mode is a two-photon decay.

\item{$\bullet$} Another possibility is that the PNGB mainly decays
into particles in an invisible sector. In this case, the event
contains a single hard photon and missing energy.

\item{$\bullet$} If \pa\ gets its mass from effective four-fermion
couplings due to extended technicolor interactions, then it
could decay to a fermion/anti-fermion pair via
two-technifermion/two-fermion couplings.  In many models,
the coupling between \pa\ and
the fermions is proportional to the fermion mass \esetc.

\item{$\bullet$} If some technifermions are colored, a color-neutral
PNGB can decay to a pair of gluons through a triangle diagram with
internal technifermions.

\item{$\bullet$} Any PNGB produced by the process $Z \to \gamma \pas$
can decay through a photon and off-shell $Z$. This three-body mode is
significant only when no two-body decay paths are open \ransim.  For
this reason, and because there is no data suggestive of this decay
path \ranrius , we do not consider this mode further.

\medskip

As we examine the LEP data for signs that PNGBs are
being produced  at the $Z$ pole, we will study the final
states that correspond to each of the two-body decay modes.  In this
way, we will cover a range of scenarios for the dominant decay
modes of the PNGB.

\newsec{Final State Kinematics}

To search the data for evidence of a light PNGB, we must
determine how the experimental cuts would affect the size of the
signal. Hence, we need to find the energies and directions of the
various final state particles.  We can divide the decay process into
two steps: the two-body decay $Z \to \pas \gamma$ and the subsequent
decay of \pa .

However \pa\ decays, the final state at LEP always contains one
photon of energy
\eqn\egam{E_\gamma = {M_Z^2 - \mp^2 \over {2 M_Z}}\ .}
For example, if $M_P$ is less than 65 GeV then $E_\gamma$ always
exceeds 20 GeV.  If the \pa\ decays invisibly, no further kinematic
information is required.

If the \pa\ decays to pairs of photons, gluons, or fermions, we
will need the energies and directions of the decay products.  We will
consider $P^a \rightarrow b{\overline b}$ explicitly.  The results may
be applied to other final states by substituting the
appropriate mass for $m_b$ in the final formulae.

Starting in the \pa\ rest frame, where the b-quarks are produced
back-to-back, the four-momenta for the $b$ and $\bar b$ quarks are,
respectively,
\eqn\grpb{\eqalign{{\cal P}_b &= ({1\over 2}M_P,{\rm
p_2}\cos\theta,{\rm
  p_2}\sin\theta\cos\omega,{\rm p_2}\sin\theta\sin\omega ) \cr
  {\cal P}_{\overline b} &= ({1\over 2}M_P,-{\rm p_2}\cos\theta,-{\rm
  p_2}\sin\theta\cos\omega,-{\rm p_2}\sin\theta\sin\omega ),}}
where
\eqn\grpp{{\rm p_2}= \sqrt{{1\over 4}M^2_P - M^2_b},}
while $\theta$ and $\omega$ are defined such that if the \pa\
boost is in the $(1,0,0)\ $ direction, the b-quark moves along
$(\cos\theta,\sin\theta\cos\omega,\sin\theta,\sin\omega)$.

Boosting to the lab frame with the beamline in the $\hat{x}$ direction
and the \pa\ three-momentum pointing along
$(\cos\alpha,\sin\alpha\cos\beta,\sin\alpha,\sin\beta)\ $, we get:
\eqn\greb{\eqalign{{\rm E}_b  &= {M_Z\over 4} \left[(1 +
\sqrt{(1-y)}\cos\theta)  + x(1 -
\sqrt{(1-y)}\cos\theta)\right] \cr
{\rm E}_{\overline b} &= {M_Z\over 4} \left[(1 -
\sqrt{(1-y)}\cos\theta)  + x(1 +
\sqrt{(1-y)}\cos\theta)\right]}}
where $y = 4M^2_b/M^2_P $ and $x = M^2_P/M^2_Z$.

To apply experimental cuts to our signal, we will need the direction
cosines of the particle momenta and the angles of separation between
the various particles.  The direction cosine between the beamline and
the b-quark is
\eqn\grbcos{\cos{\cal X}_2 = {{[(1 +
\sqrt{(1-y)}\cos\theta) - x(1 - \sqrt{(1-y)}\cos\theta)]\cos\alpha -
2\sqrt{x}\sin\theta\cos\omega\sin\alpha}\over
{\sqrt{{[(1 + \sqrt{(1-y)}\cos\theta) - x(1 -
    \sqrt{(1-y)}\cos\theta)]}^2 + 4x\sin^2\theta}}}}
The direction cosine $\cos{\cal X}_3$ for the ${\overline b}$-quark is
obtained by reversing the sign of each $\cos\theta$ and $\sin\theta$ in
equation \grbcos .
The direction cosine between the photon and the b-quark is
\eqn\grbgamcos{\cos{\cal X}_{12} = {{-[(1 +
\sqrt{(1-y)}\cos\theta) - x(1 - \sqrt{(1-y)}\cos\theta)]}\over{\sqrt{{[(1 +
\sqrt{(1-y)}\cos\theta) - x(1 - \sqrt{(1-y)}\cos\theta)]}^2 +
4x\sin^2\theta}}}}
The direction cosine $\cos{\cal X}_{13}$ between the photon and the
anti-b-quark follows by sending $\cos\theta \to -\cos\theta$ in
equation \grbgamcos .
The angle between the quark and anti-quark is
\eqn\grang{{\cal X}_{23} = 2\pi - ({\cal X}_{12} + {\cal X}_{13})}
since the event is planar.

\newsec{$Z \to \gamma\pas \to \gamma\gamma\gamma$}

The three photon events should yield a distinctive signal,
including a hard photon whose energy is fixed by $M_P$.  We used a
Monte Carlo integration over $\theta$, $\alpha\,$ and $\omega\,$ to
calculate the effects of the DELPHI experimental cuts \grdel\ on the
signal:

\medskip

\item{$\bullet$} {\tt The energy of each photon is greater than
10 GeV.}  This eliminates the signal for \pa\ heavier than 80 GeV.

\item{$\bullet$} {\tt All three photons in a given event should have
polar angles  between $20^\circ$ and $160^\circ$; two of the three
should lie in the restricted range $42^\circ < \theta < 138^\circ$.}
This uniformly removes about 20\% of the signal.

\item{$\bullet$} {\tt The most energetic photon should carry energy
greater than 20 GeV.} This has no effect on the number of signal events.

\item{$\bullet$} {\tt The most energetic photon should have polar
angle between $40^\circ$ and $140^\circ$.}\hfil\break
Like the second cut, this one removes 20\% of the signal.

\item{$\bullet$} {\tt The angular separations between the least
energetic photon and each of \hfil\break
the other two should be greater
than $20^\circ$.} This cut mainly affects light \pa\ which are produced
with large momenta and therefore decay into two photons with small
angular separation.

\medskip

\noindent{In addition, the DELPHI experiment \grdel\ required the absence of
vertex detector tracks pointing to the clusters in the electromagnetic
calorimeter (`photons') and the general absence of energetic tracks
coming from the beam-crossing point.}

Depending on the PNGB mass,
between 40\% and 75 \% of the signal survives the cuts on the photons,
as shown in fig. 1.

The size of the expected signal may now be compared with the
experimental upper limit \grdel\ on the branching ratio:
B.R.($Z\to\gamma\gamma\gamma) < 1.7 \times 10^{-5}$.  We set the PNGB
decay constant to $f = 123$ GeV, and used
\zdecay\ to find the maximum value of $[N_{TC} {\cal A}_{Z\gamma}]$
as a function of $M_P$, as shown in fig. 2. The limit is of order 2-3
for masses below about 60 GeV, and weakens with increasing
mass.  The lowest mass for which the bounds apply is set by the angular
separation cut; the highest mass, by the 10 GeV photon energy cut.

\medskip

%
$$
\epsfxsize= 4.5truein \epsfbox[21 145 592 600]{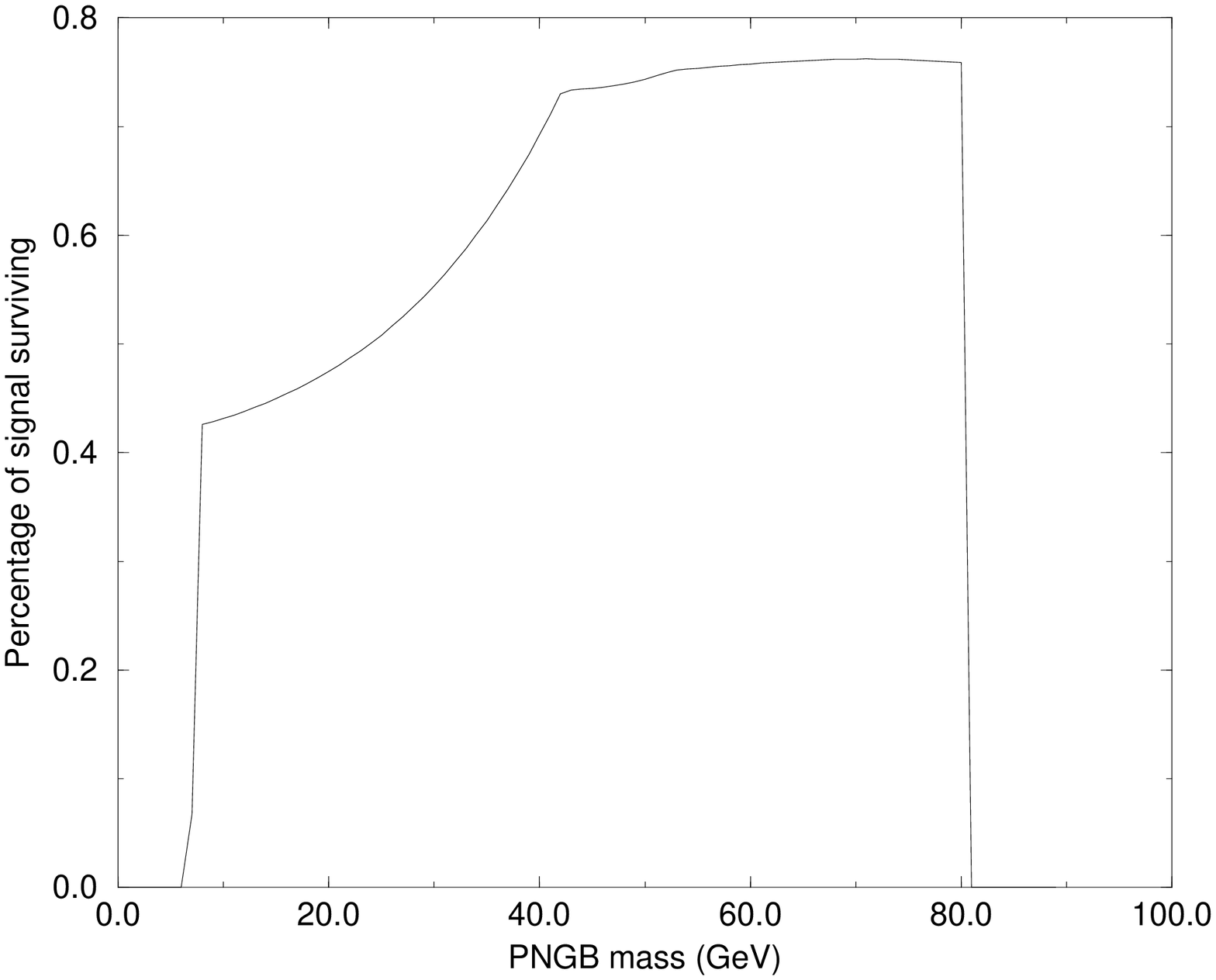}
$$
\vbox{\centerline{Fig. 1. Percentage of $Z\to \gamma\pas \to
\gamma\gamma\gamma$ signal surviving cuts}
\vskip -0.08in \centerline{used by the DELPHI experiment,
as a function of PNGB mass.}
\vskip -0.08in \centerline{The left boundary comes from the
angular separation cut; the shape at}
\vskip -0.08in \centerline{low mass and the right-hand boundary, from
the 10 GeV photon energy cut.}}

%
$$
\epsfxsize=4.5truein \epsfbox[21 145 592 600]{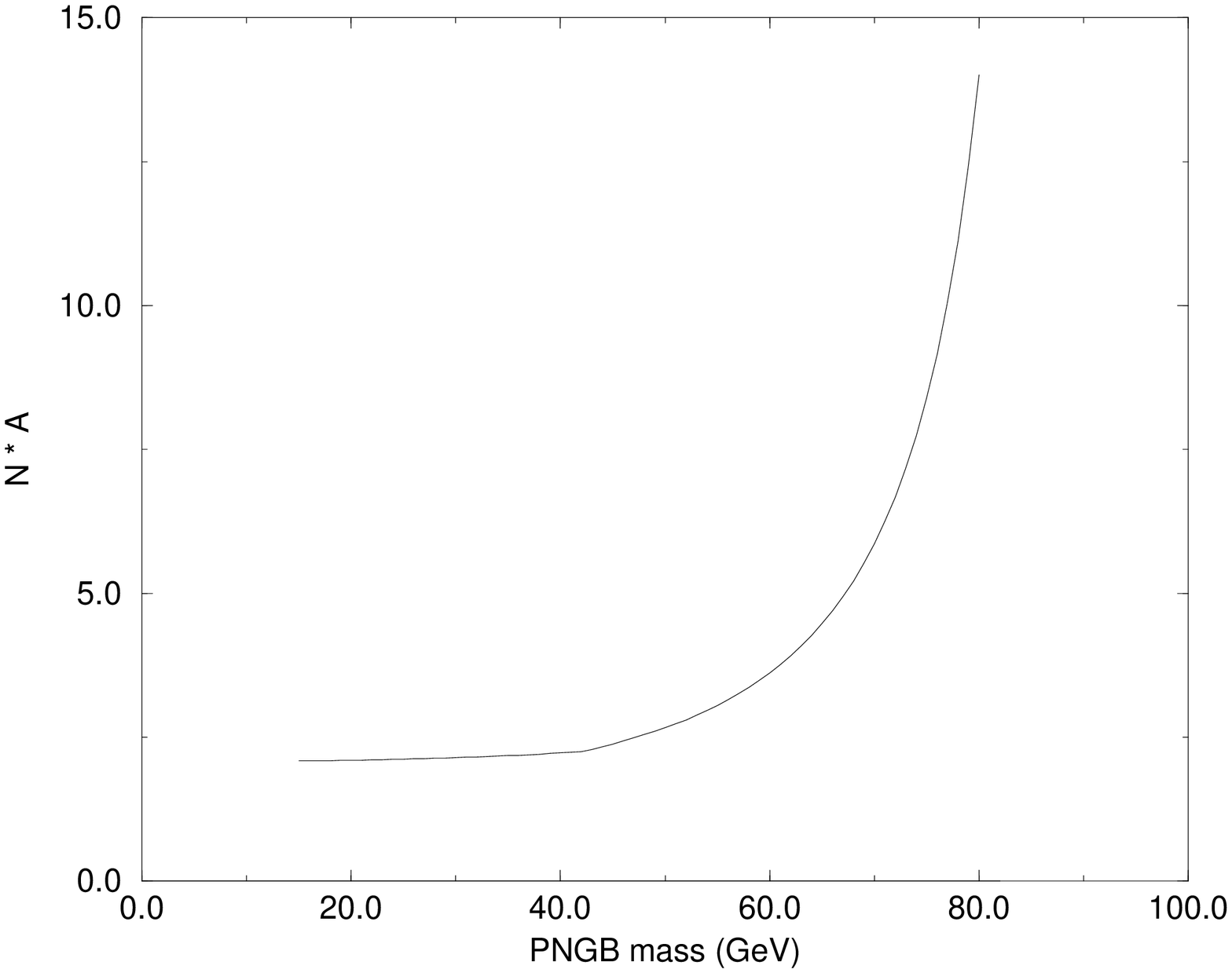}
$$
\vbox{\centerline{Fig. 2. Upper bound (at 95\% c.l.) on $N_{TC}
{\cal A}_{Z\gamma}$ as a function of \pa\ mass}
\vskip -0.08in \centerline{in models where \pa\ decays
dominantly to photons and $f = 123$ GeV.}
\vskip -0.08in\centerline{Derived from eq. (2.5)
and DELPHI data on $Z\to \gamma\gamma\gamma$.}}

\newsec{$Z \to \gamma\pas \to \gamma + E\!\!\!/ $}

If the scalar's dominant decay is invisible, only the photon from the
$Z$ decay  will be directly observeable.  The experimental candidate
events are characterized by one energetic photon in an otherwise
empty detector: no tracks, muon hits, or other energy deposits
(except as consistent with noise).  To suppress background to new
physics, cuts are also placed on the single photon.  To assess
the effects of these on the signal, we performed a Monte
Carlo integration over the variable $\theta$, imposing the following
conditions \grlthrephot\ used by the L3 Collaboration:

\medskip

\item{$\bullet$} {\tt The energy of the photon should be greater
than 15 GeV.}  This eliminates the signal for PNGB's heavier than 75
GeV.

\item{$\bullet$} {\tt The photon must have a polar angle between
$20^\circ$ and $160^\circ$ (excluding the \hfil\break
regions
$34.5^\circ - 44.5^\circ$ and $135.5^\circ - 145.5^\circ$).}  Since
the $Z$ decay to scalar plus photon is isotropic, this cut uniformly
reduces the signal by approximately one third.

\medskip

The L3 paper \grlthrephot\ plots the number of events as a function of
photon energy, using energy bins 2 GeV wide.  No bin shows an excess of
more than one event over expected standard model backgrounds from
$\nu\bar\nu\gamma$, $e^+ e^- \gamma$ and $\gamma\gamma\gamma$ final
states.  Assuming $f = 123$ GeV and applying Poisson statistics, we
find at 95\%{\it c.l.} that $[N_{TC} {\cal A}_{Z\gamma}] \lta 1.5$ for
$40 < M_P < 70$ GeV.  The precise limit as a function of $M_P$ is
shown in Figure 3.

We performed a similar analysis based on the OPAL data \gropalphot .
In this experiment, single-photon events were selected if the photon
carried energy above 1.75 GeV and restricted to the polar region
$|\cos\alpha | < 0.7 $.  OPAL reports \gropalphot\ an upper
bound (at 95\% {\it c.l.}) of $4.3 \times 10^{-6}$ on the branching
ratio for radiative decay of the $Z$ to an invisibly-decaying scalar
particle $X$ with mass less than 64 GeV, and an upper bound of $1.4
\times 10^{-5}$ for a mass less than 84 GeV.  This
sets the approximate limit $[N_{TC} {\cal A}_{Z\gamma}] < 1$ for
PNGB masses less than 40 GeV.  As shown in fig. 3, the bounds from
the OPAL data are weaker than those from L3 for values of $M_P$
accessible to both experiments.

\medskip
%
$$
\epsfxsize=4.5truein \epsfbox{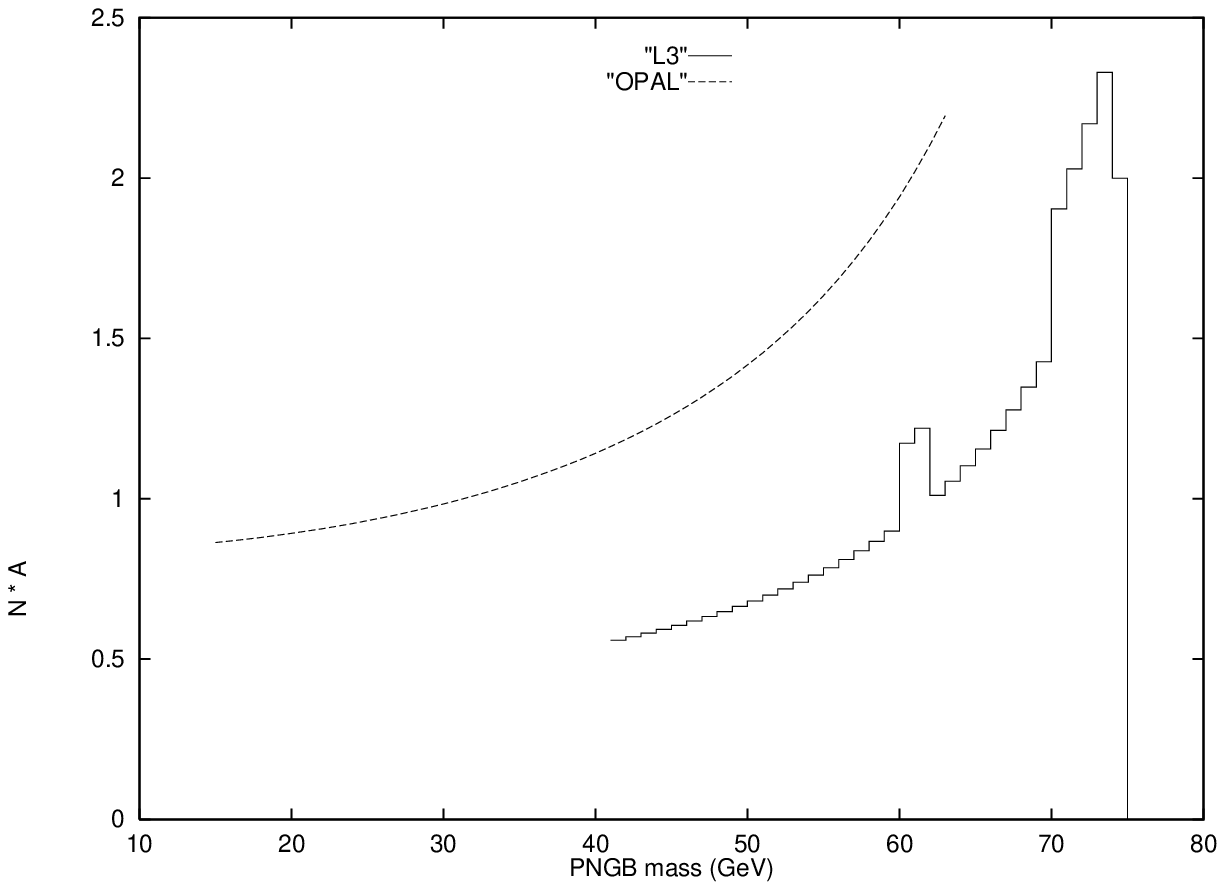}
$$
\vbox{\centerline{Figure 3. Upper bound (at 95\% c.l.) on $N_{TC}
{\cal A}_{Z\gamma}$ as a function.}
\vskip -0.08in \centerline{of PNGB mass in models where PNGB decays
invisibly and $f = 123$ GeV.}
\vskip -0.08in\centerline{Derived from eq. (2.5)
and L3 and OPAL data on $Z\to \gamma + E\!\!\!/ $.}}

\newsec{$Z \to \gamma\pas \to \gamma\ {\rm jet\ jet}$}

If $P^a$ can decay into gluons, this mode dominates.  Otherwise, if
the PNGB couples directly to ordinary fermions in proportion to the
fermion masses, then $\pas \to b\bar b$ is expected to be the most
important mode, with decays to $c\bar c$ and $\tau^+\tau^-$ taking
over for $M_P < 2 m_b$.  In either case, the L3 data on isolated hard
photons in hadronic $Z$ decays \grlthrehadr , can be used to search
for \pa .  We explicitly discuss the case in which fermionic decays of
\pa\ predominate.  The limits on models in which $\pas \to g g$
dominates are nearly identical since no flavor tagging is employed in
the selection cuts and since the jet-jet angular separation cut
essentially eliminates most \pa\ lighter than $2 m_b$.

Following the L3 experimental analysis \grlthrehadr, we applied the
following cuts to our signal:

\medskip

\item{$\bullet$}{\tt The photon energy is greater than 5 GeV.}  This
suppresses the signal for \pa\ heavier than 85 GeV.

\item{$\bullet$}{\tt The photon should lie between polar angle $45^\circ$ and
$135^\circ$.}  This uniformly reduces the signal strength by about
$30\%$.

\item{$\bullet$}{\tt The hadronic jets should lie in the region $5^\circ <
\theta <  175^\circ$.} This reduces the signal by
another few percent.

\item{$\bullet$}{\tt The jets should be separated by at least
$20^\circ$ from one another and from \hfil\break
the photon.}  The main
effect is to suppress the signal for low scalar masses.

\medskip

\noindent{About 70\% of the signal survives these cuts for $P^a\ $
masses between 8 and 85 GeV.}

The L3 collaboration \grlthrehadr\ reports a
limit on radiative $Z$ decay to a hadronically decaying
narrow resonance, $Y$, that may be approximately written
as
\eqn\limlte{\sigma(e^+e^- \to Z \to \gamma Y) \cdot {\rm B.R.}(Y\to
hadrons) \lta 5\ {\rm pb}, \ \ \ \ {\rm for}\ \  M_Y \lta 85\ {\rm
GeV}.} Comparison with our expected
signal for a hadronically decaying \pa\ implies
$[N_{TC}\ {\cal A}_{Z\gamma}] \leq 5$
for $M_P \lta 50$\ GeV and $f = 123$ GeV.  The bound loosens
to $[N_{TC} {\cal A}_{Z\gamma}] \leq 10$ for $M_P
\approx 60$ GeV and becomes even weaker for heavier $P^a$
because the production rate declines so steeply
with increasing $M_P$.

\newsec{Applications to Technicolor Models}

Having reported the bounds that LEP sets on neutral
PNGB in the general form of limits on
$[N_{TC} {\cal A}_{Z\gamma}]$, we
now apply them to several technicolor scenarios.  This
allows us to see what classes of models are most strongly
constrained now and which will be probed by more
stringent experimental limits on rare $Z$ decays.

The benchmark technicolor model which includes PNGB is
one-family technicolor \bench .  In this model, the technipion decay
constant is $f = v/2 = 123$ GeV, and the neutral PNGB
can be described in terms of technifermion quantum numbers as
\eqn\pngbbench{P^1 \sim (\bar Q \gamma_5 Q - 3 \bar L \gamma_5 L)\ ,
\ \ \ \ \ \ \ \  P^3 \sim (\bar Q\gamma_5 \tau_3 Q - 3 \bar L \gamma_5
\tau_3 L) }
where the superscript on $P$ indicates the dimension of
the custodial isospin representation. These can decay to two jets
either through direct ETC couplings to quarks or through the coupling
of gluons to the techniquarks. The corresponding anomaly factors are
\eqn\anombench{
{\cal A}^1_{Z\gamma} = {1\over 3\sqrt3}s^2_W \approx 0.044\ ,\ \ \ \
\ \ \ \ {\cal A}^3_{Z\gamma} = {1\over \sqrt3}({1\over4} - s^2_W)
\approx 0.012\ .}
Note that, as mentioned in section 2, the state
with I=1 has an anomaly factor suppressed by
$({1\over4} - s^2_W)$.   Both anomaly factors are so small that the
LEP limits on radiative hadronic $Z$ decays set no useful bounds on
one-family technicolor.

Since $\Gamma(Z\to \gamma P)$ goes as $(N_{TC}
{\cal A}_{Z\gamma} / f)^2$, models more likely to be
probed by the LEP data will need to have
smaller $f$ or larger ${\cal A}_{Z\gamma}$ than
one-family technicolor.  Several models with small
technipion decay constants have recently been proposed;
we consider these first.

In ref. \appter , the authors introduce
one-family technicolor models in which the QCD
interactions combine with near-critical ETC interactions
to enhance the techniquark masses relative to those of
the technileptons.  The main attraction of such models
is that non-degenerate technifermions can potentially
lead to a small value of the electroweak radiative
correction parameter, $S$ \precision .  The side-effect of interest
here is that the technipions composed of technineutrinos and
technileptons can be relatively light -- with masses as
low as 50-100 GeV -- and have small decay constants (of
order 40 GeV).  Due the splitting between the
techniquarks and technileptons, the model has a smaller chiral flavor
symmetry than one-family technicolor, and the neutral PNGB are
\eqn\pngbat{P^3_Q \sim \bar Q \gamma_5 \tau^3 Q\ ,\ \ \ \ \ \
P^3_L \sim \bar L \gamma_5 \tau^3 L\ ,\ \ \ \ \ \
P^1 \sim \bar Q \gamma_5 Q - 3 \bar L \gamma_5 L\ .}
The NGB that is eaten by the $Z$ is largely composed of
techniquarks ($f_Q >> f_L$) and the remaining PNGB mass
eigenstates are approximately (in the limit of large
isospin breaking)
\eqn\pngbatm{P_N \sim \bar N \gamma_5 N\ ,\ \ \ \ \ \ \ \
P_E \sim \bar E \gamma_5 E\ .}
The lighter of these, $P_N$, has ${\cal A}_{Z\gamma} = 0$
since the technineutrino carries no electric charge.  The
heavier $P_E$ can still be light enough to be produced at
LEP.  Although $P_E$ has an $I=0$ component,
the terms in its anomaly factor proportional to $-2s^2_W$
and to $1 - 2s^2_W$ have identical coefficients, so the
full anomaly factor is
\eqn\anomat{{\cal A}^E_{Z\gamma} = {1\over\sqrt{2}}({1\over4} -
s^2_W)\ .}
Even the reduced $f_L$ in this model is not sufficient to
overcome this.  In consequence,
$P_E$ will not be visible at LEP no matter how it decays.

The multiscale TC models introduced in \eichlan\ include
technifermions in large technicolor representations and can also have
light neutral PNGB with small decay constants \eichlan\lanram\lane.
While the lightest PNGB can\foot{Based on the results of \lubicztwo\
and lowering the scaling factor, $\kappa$, to 1.0 \eichlan .} have a
mass below $M_Z$, this PNGB, however, is largely composed of the I=1
state $P^3_L \sim \bar L\gamma_5 \tau_3 L$ so that although $f_L \sim
30$ GeV, the anomaly factor is suppressed by $({1\over4} - s^2_W)$.
In the most optimistic case, in which the PNGB decays half the time to
a pair of photons \lubicztwo, the model with $N_{TC} = 6$ gives a
signal 10 times smaller than the LEP bound.  If the decays are
primarily hadronic (through ETC interactions), the signal lies even
further below the relevant limit. P.

Clearly, small $f$ is not enough: we also need models with large
${\cal A}_{Z\gamma}$, e.g., those in which the
lightest technifermions belong to larger representations
of $SU(2)_W$.  Consider, for example, a scenario
suggested in \manran\ in which the left-handed
technifermions are a weak isotriplet of techniquarks, Q,
with hypercharge $y$ and one of technileptons, L, with
hypercharge $-3y$, while the right-handed technifermions
are weak singlets.  In this case, $f = v/4$ and there
are I=0 and I=2 states
\eqn\pngbmr{P^1 \sim (\bar Q \gamma_5 Q - 3 \bar L \gamma_5
L)\ ,\ \ \ \ \ \ \ \
P^5_+ \sim (\bar Q \gamma_5 T^8 Q + \bar L \gamma_5 T^8
L) \ ,}
where $T^8 = diag(1,-2,1)$, with sizeable anomaly
factors:
\eqn\anommr{{\cal A}^1_{Z\gamma} = 6 \sqrt{2} s^2_W y^2\ ,\ \ \ \ \ \
\ \  {\cal A}^5_{Z\gamma} = {1\over\sqrt{3}}(1 - 2 s^2_W)\
.}
No matter what the (model-dependent) dominant decays of these
technipions, the LEP data place strong limits on the size of $N_{TC}$
as a function of $M_P$.

If the PNGB decay invisibly, with $f = v/4$ the experimental limit is
(cf. section 5)
\eqn\liminv{ 2 N_{TC} {\cal A}_{Z\gamma}\ \leq\
1\ \ (2) \ \ \ \ \ {\rm  when}\ \ \ M_{P}\ \lta\ 65\ {\rm GeV}\ \ (75\
{\rm GeV})\ .}
If the I=2 state is sufficiently light, then \liminv\ combined with
\anommr\ yields
\eqn\nliminv{N_{TC}\ \leq\ 1\ \ (3) \ \ \ \ {\rm
when}\ \ \ M_{P^5_+}\ \lta\ 65\ {\rm GeV}\ \ (75\ {\rm GeV})\ .}
so that LEP has excluded all such models with $M_{P^5_+} \lta 65$ GeV
and allows only $N_{TC} = 2, 3$ if $65 \lta M_{P^5_+} \lta 75$ GeV.  If
only the I=0 state is light enough to be produced at LEP, then for
techniquark hypercharge $y > 1/3$, the limits are
equivalent to those shown above.  The non-observation of
these PNGB therefore implies that either $P^1$ and $P^5_+$
are heavy or $y$ is small.

If $\gamma\gamma$ decays dominate, the LEP data tell us (cf. section 4)
\eqn\limgg{ 2 N_{TC} {\cal A}_{Z\gamma} \leq 2.5 \ \ (5)\ \ \ \ {\rm
for}\ \ \ M_{P} \lta 50 {\rm GeV}\ \ (70\ {\rm GeV})\ .}
This excludes models in which $P^1$ has a mass less than 50
GeV and the techniquark hypercharge exceeds $1\over2$.  Models in
which $P^5_+$ is light or the techniquark hypercharge is smaller
are allowed within the following constraints
\eqn\nlimgg{\eqalign{ N_{TC} \leq 4 \ \ \ \ &{\rm
for}\ \ \ M_{P^5_+} \lta 50 {\rm GeV}\cr
N_{TC} \leq 5 \ \ \ \ &{\rm
for}\ \ \ M_{P^1} \lta 50 {\rm GeV}\ {\rm and}\ y \geq {1\over3}\ .}}
The extension to heavier PNGB and smaller $y$ follows from \anommr\
and \limgg .

Finally, if hadronic decays of $P^1$ or
$P^5_+$ dominate, the data \limlte\ place an upper bound of
5 pb on the quantity $\left[ 70\% \cdot \sigma^Z_{tot} \cdot B.R.(Z \to
\gamma P^a)\right]$ where the B.R. comes from \zdecay\ and the factor
of 70\% is the roughly mass-independent acceptance of the cuts.  In the
range $20 < M_P < 60$ GeV, this quantity is approximately linear in
$M_P$:
\eqn\aplin{{1\over4} \left[ 5.5 + 2.2 (1 - {M_P \over {20 {\rm
GeV}}})\right] {\rm pb}\ \cdot\ (N_{TC} {\cal A}_{Z\gamma})^2 }
where $f = v/4$.  Eq. \aplin\ does not set a strong lower limit on the
mass of $P^5_+$; even if $M_P{^5}$ is as small as 20 GeV, $N_{TC} = 6$
is allowed.  On the other hand, for $y=1$, the $P^1$ must weigh at
least 60 GeV; if $y = {1\over2}$, then $M_{P^1} > 40$ GeV unless
$N_{TC} \leq 5$.

\newsec{Summary}

LEP data on rare $Z$ decays are beginning to set limits on the
presence of the light neutral pseudo Nambu-Goldstone bosons
characteristic of non-minimal technicolor models.  Not surprisingly,
the data most strongly constrains models with a large $Z \gamma \pas$
coupling.  Probing multiscale technicolor models which have a smaller
$Z\gamma\pas$ coupling but also have small technipion decay constants
will be possible with an improvement of order a factor of 10 in the LEP
bounds on rare $Z$ decays.

\bigbreak\bigskip\bigskip\centerline{{\bf Acknowledgments}}\nobreak
The authors thank Bing Zhou for experimental information. We also thank
Sekhar Chivukula and Ken Lane for comments on the manuscript. EHS
acknowledges the support of an NSF Faculty Early Career Development
(CAREER) Award and the hospitality of the Aspen Center for Physics
where this work was completed. {\it This work was supported in part by
the National Science Foundation under grant PHY-9501249 and by the U.S.
Department of Energy under grant DE-FG02-91ER40676.}

\listrefs

\bye